\documentclass[osajnl,twocolumn,showpacs,superscriptaddress,10.5pt]{revtex4-1} 
\usepackage{amsmath,amssymb,graphicx}
\begin{document}

\title{Laser frequency locking with 46 GHz offset using an electro-optic modulator\\
for magneto-optical trapping of francium atoms}

\author{K. Harada}\email{kenichi.harada.a1@tohoku.ac.jp}
\affiliation{Cyclotron and Radioisotope Center (CYRIC), Tohoku University, 6-3 Aramaki-aza Aoba, Aoba-ku, Miyagi 980-8578, Japan}
\author{T. Aoki}
\affiliation{Graduate School of Arts and Sciences, University of Tokyo, 3-8-1 Komaba, Meguro-ku, Tokyo 153-8902, Japan}
\author{S. Ezure}
\author{K. Kato}
\author{T. Hayamizu}
\affiliation{Cyclotron and Radioisotope Center (CYRIC), Tohoku University, 6-3 Aramaki-aza Aoba, Aoba-ku, Miyagi 980-8578, Japan}
\author{H. Kawamura}
\author{T. Inoue}
\affiliation{Frontier Research Institute for Interdisciplinary Sciences, Tohoku University, 6-3 Aramaki-aza Aoba, Aoba-ku, Miyagi 980-8578, Japan}
\affiliation{Cyclotron and Radioisotope Center (CYRIC), Tohoku University, 6-3 Aramaki-aza Aoba, Aoba-ku, Miyagi 980-8578, Japan}
\author{H. Arikawa}
\author{\\T. Ishikawa}
\author{T. Aoki}
\author{A. Uchiyama}
\author{K. Sakamoto}
\author{S. Ito}
\author{M. Itoh}
\author{S. Ando}
\affiliation{Cyclotron and Radioisotope Center (CYRIC), Tohoku University, 6-3 Aramaki-aza Aoba, Aoba-ku, Miyagi 980-8578, Japan}
\author{A. Hatakeyama}
\affiliation{Department of Applied Physics, Tokyo University of Agriculture and Technology, 2-24-16 Nakacho, Koganei, Tokyo 184-8588, Japan}
\author{K. Hatanaka}
\affiliation{Research Center for Nuclear Physics, Osaka University, 10-1 Mihogaoka, Ibaraki, Osaka 567-0047, Japan}
\author{K. Imai}
\affiliation{Advanced Science Research Center, Japan Atomic Energy Agency, 2-4 Ooaza Shirakata, Tokai, Naka, Ibaraki 319-1195, Japan}
\author{T. Murakami}
\affiliation{Department of Physics, Kyoto University, Kitashirakawa-Oiwakecho, Sakyo-ku, Kyoto 606-8502, Japan}
\author{H. S. Nataraj}
\affiliation{Indian Institute of Technology Roorkee, Roorkee, Uttarakhand 247667, India}
\author{Y. Shimizu}
\affiliation{Department of Physics, Tohoku University, 6-3 Aramaki-aza Aoba, Aoba-ku, Miyagi 980-8578, Japan}
\author{T. Sato}
\affiliation{Department of Physics, Tokyo Institute of Technology, 2-12-1 O-okayama, Meguro, Tokyo 152-8551, Japan}
\author{T. Wakasa}
\affiliation{Department of Physics, Kyushu University, 6-10-1 Hakozaki, Higashi, Fukuoka 812-8581, Japan}
\author{H. P. Yoshida}
\affiliation{Research Center for Nuclear Physics, Osaka University, 10-1 Mihogaoka, Ibaraki, Osaka 567-0047, Japan}
\author{Y. Sakemi}
\affiliation{Cyclotron and Radioisotope Center (CYRIC), Tohoku University, 6-3 Aramaki-aza Aoba, Aoba-ku, Miyagi 980-8578, Japan}

\begin{abstract}
We demonstrated a frequency offset locking between two laser sources using a waveguide-type electro-optic modulator (EOM) with 10th-order sidebands for magneto-optical trapping of Fr atoms. The frequency locking error signal was successfully obtained by performing delayed self-homodyne detection of the beat signal between the repumping frequency and the 10th-order sideband component of the trapping light. Sweeping the trapping-light and repumping-light frequencies with keeping its frequency difference of 46 GHz was confirmed over 1 GHz by monitoring the Doppler absorption profile of I$_2$. This technique enables us to search for a resonance frequency of magneto-optical trapping of Fr.
\end{abstract}

\maketitle

\section{Introduction}
The francium atom, which is the heaviest alkali element, has advantages as regards examination of the electron electric dipole moment (EDM) \cite{Lu} and measurements of atomic parity nonconservation (APNC) \cite{Simsarian1, Atutov}. The Fr atom exhibits high signal sensitivity because of the large effective electric field due to the large nuclear charge. The enhancement factor of a Fr atom for electron EDM is approximately 900 \cite{Roberts, Mukherjee}, and the E1 amplitude for a nuclear-spin-independent APNC in a Fr atom is approximately 17 times larger than that of a Cs atom \cite{Roberts}. The finite value of the electron EDM serves as the signature of violation of both parity ($P$) and time-reversal ($T$) invariance and APNC provide the signature of $P$ violation. Searches for the violations of fundamental symmetries are important for evaluating a variety of the new physics beyond the Standard Model \cite{Pospelov}.

In conventional experiments using atomic and molecular beams \cite{Wood, Regan, Hudson, ACME}, systematic errors and short interaction time of the atoms/molecules with the applied electric field are major obstacles restricting the measurement sensitivity. Laser-cooled and trapped atoms elongate an interaction time between the atoms and applied electric field \cite{Chin}. Therefore it is expected to improve the measurement sensitivity of the EDM and APNC. Moreover, this method reduces the atomic velocity and confines the atoms within a small region, which suppresses several systematic errors. At present, we are conducting a high-sensitivity search for the electron EDM using laser-cooled Fr atoms \cite{Sakemi, Kawamura, Inoue, Harada}. Radioactive Fr is produced by a nuclear fusion reaction in a heated Au target bombarded by a beam of accelerated 100 MeV $^{18}$O$^{5+}$ at the Cyclotron and Radioisotope Center. The maximum yield of Fr beam is estimated to be about 10$^6$ ions/s. Fr has also been produced at other facilities using different techniques \cite{Lu, Simsarian1, Atutov, Collister}.

To stabilize the laser frequency of a magneto-optical trap (MOT), the frequency reference, such as signals obtained by saturated absorption spectroscopy of atoms, is generally required for the frequency stabilization of the trapping light. It is also typically necessary for the repumping light in order to stabilize its frequency. However, Fr atomic vapor cells as frequency references are not available because no stable Fr isotopes exist. To search for the resonance frequency, the frequency of trapping light was scanned over a 60 MHz range at 0.5 MHz/s by monitoring the fluorescence of Fr and the Doppler-broadened absorption signal of I$_2$ for first MOT of Fr \cite{Simsarian1}. The frequency of repumping light is scanned over 500 MHz at a rate of 4 kHz \cite{Simsarian2}. In these experiments, the frequencies of trapping and repumping lights are scanned independently. By contrast, the frequency of trapping (repumping) light in Fr experiment is stabilized by using the scanned Fabry-P\'erot cavity. The length of cavity is scanned to monitor and keep the frequency difference between the trapping (repumping) light and I$_2$-stabilized He-Ne laser \cite{Atutov, Zhao}. The cavity with transfer laser is also utilized to determine the frequency difference \cite{Sanguinetti}. However, the transfer cavity technique is not useful to search for a resonance frequency of MOT of Fr by sweeping the trapping-light and repumping-light frequency with keeping its frequency difference of 46.1 GHz.

\begin{figure}
\centerline{\includegraphics[width=0.9\columnwidth]{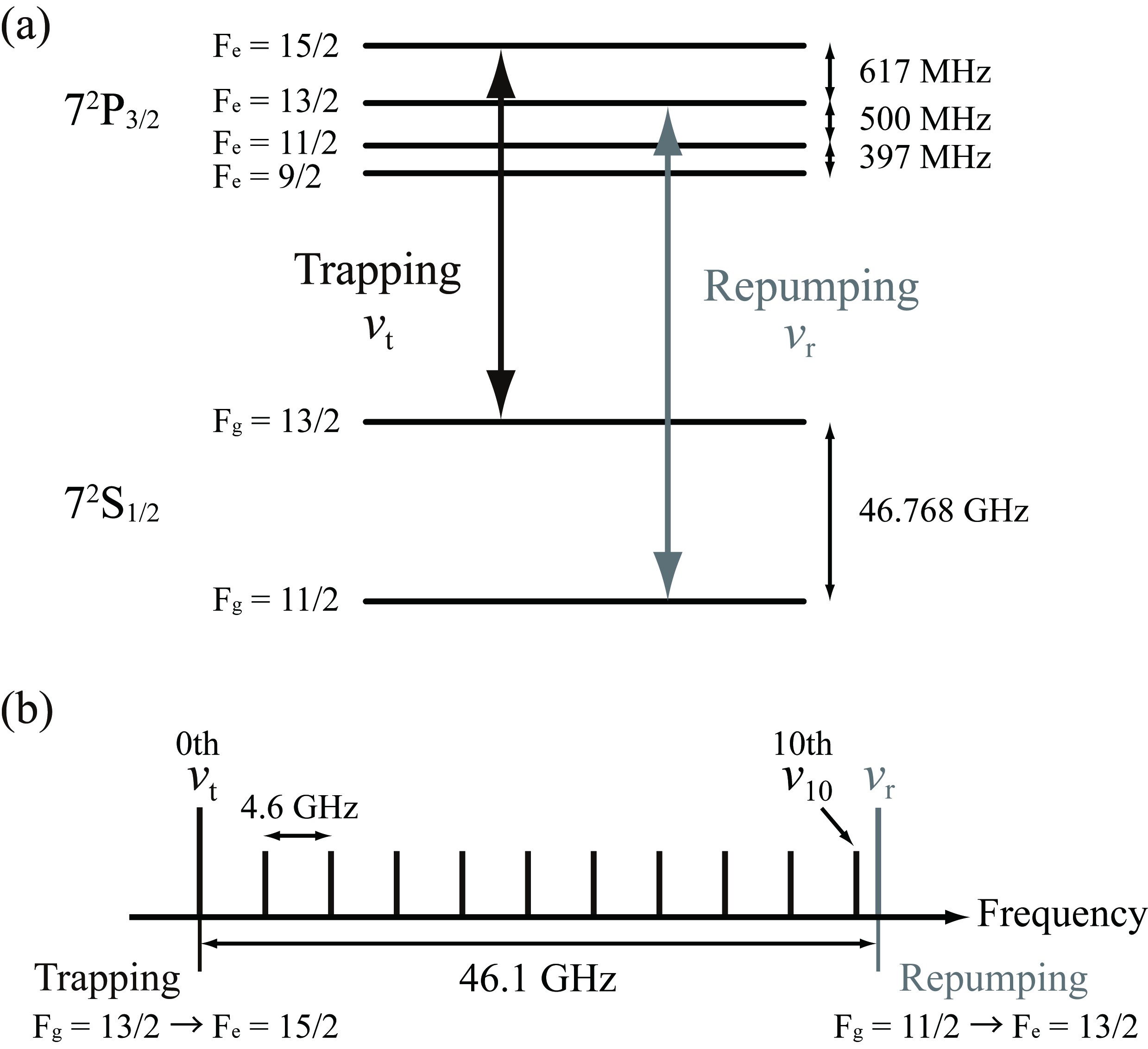}}
\caption{(a) Energy diagram of D2 transitions of $^{210}$Fr atom. The black (gray) arrow represents the transition of trapping (repumping) light. (b) Frequency relation and laser frequency. $\nu_\textrm{t}$ is the trapping frequency, $\nu_{10}$ is the 10th sideband of trapping light. $\nu_\textrm{r}$ is the repumping frequency. The frequency difference between $\nu_\textrm{r}$ and $\nu_{10}$ is typically less than 1 GHz.}
\label{Energy}
\end{figure}

In this paper, we demonstrate a frequency offset locking with 46 GHz between the trapping and repumping light beams, for the magneto-optical trapping of Fr atoms using an electro-optic modulator (EOM). A frequency-offset locking technique can lock the frequency difference between the two laser sources to a constant value \cite{Schunemann}. This locking is used for frequency stabilization and also as a means of scanning both lasers simultaneously, while maintaining a constant frequency difference within a capture range. Here, the capture range is the range in which the frequency of one laser follows the frequency sweep of the other with a 46-GHz frequency difference, using a servo loop. The offset locking technique offers alternative method to stabilize the laser frequencies for Fr MOT experiment.

\section{Experiment}

\begin{figure}
\centerline{\includegraphics[width=1\columnwidth]{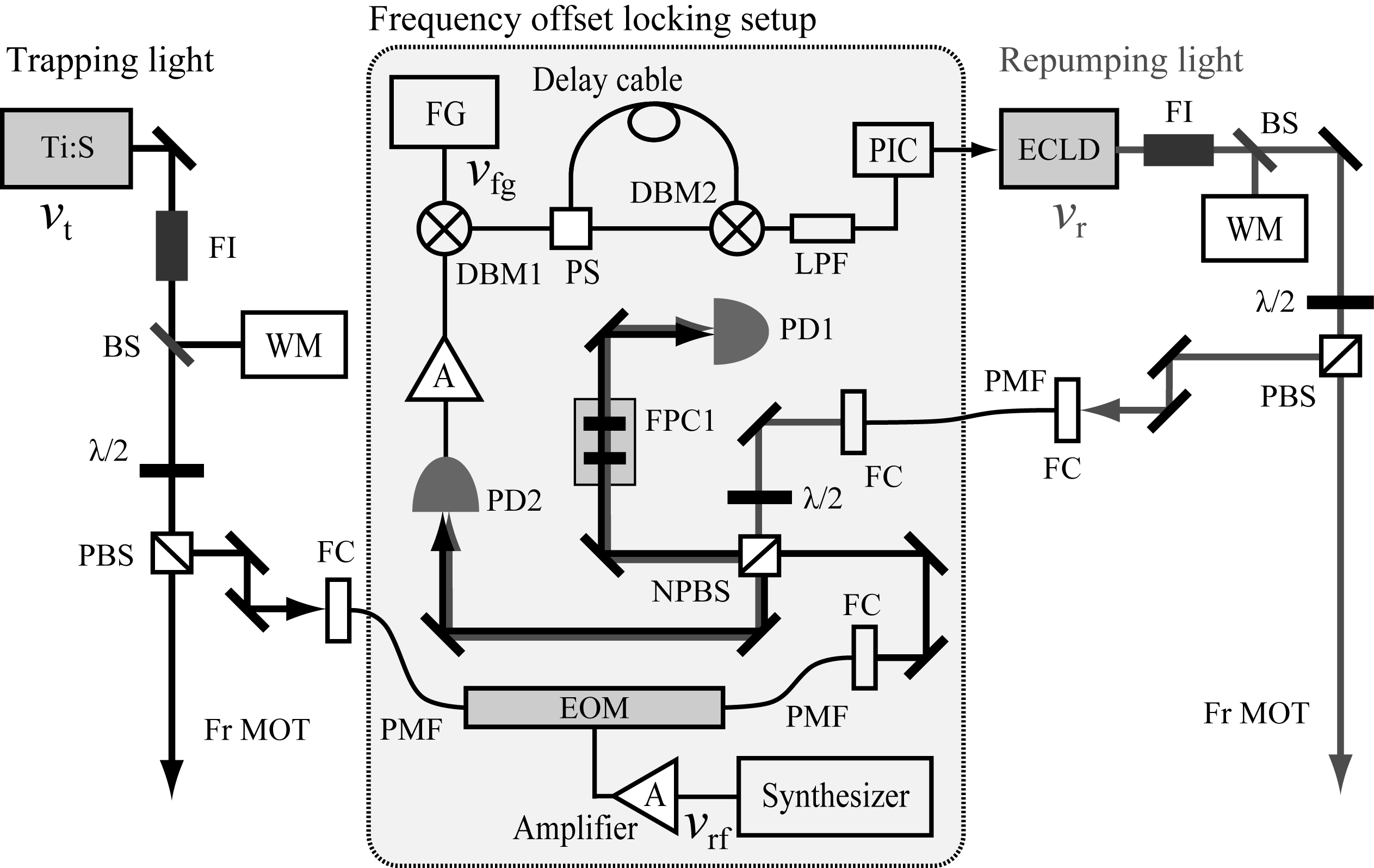}}
\caption{Schematic of frequency offset locking experimental setup.}
\label{Setup}
\end{figure}

The frequency difference between the hyperfine splittings of $7S_{1/2}$ of a $^{210}$Fr atom is 46.768 GHz, as shown in Fig. \ref{Energy}(a). The frequency difference between $7P_{3/2}, F_{\mathrm{e}} = 15/2$ and $F_{\mathrm{e}} = 13/2$ is 617 MHz \cite{Sansonettia}. Here, $F_{\mathrm{g}}$ and $F_{\mathrm{e}}$ represent the total angular momenta of the ground and excited states, respectively. Therefore, as the beat frequency between the trapping and repumping light beams is about 46.1 GHz (which includes a red-detuning of several tens of MHz for MOT), the beat frequency of 46.1 GHz has to be detected. However, it is difficult for a photodetector to detect a 46-GHz frequency directly. On the other hand, if one generates the sideband frequency with 46 GHz of trapping light and locks the frequency difference between this sideband and repumping light frequency, it is easy to lock the frequency difference between the trapping and repumping lights. However, in general, it is difficult to generate frequency components with 46 GHz at the first-order diffraction or sidebands through an acousto-optic modulator or an EOM. Here, instead of directly generating a first-order 46-GHz sideband, we use the EOM, which can generate a 46-GHz component as a 10th-order sideband by injecting the radio frequency (RF) of 4.6 GHz into the EOM, as shown in Fig. \ref{Energy}(b). The frequency difference between the repumping light and 10th-order component of the trapping light is typically less than 1 GHz, which can be easily observed as a beat signal using a standard-fast-silicon photo detector. The error signal can be obtained via the delayed self-homodyne detection of the beat signal. Stabilization of frequency difference between the repumping-light frequency and 10th sideband of trapping-light frequency causes the offset locking of 46.1 GHz. This provides a simple and low cost method of maintaining a constant frequency difference of 46.1 GHz between the two laser sources.

The experimental setup is depicted in Fig. \ref{Setup}. A single-frequency Ti:sapphire (Ti:S) laser source (Coherent, MBR-110) for the trapping light was tuned to the $7S_{1/2}$ ($F_{\mathrm{g}} = 13/2$) $-$ $7P_{3/2}$ ($F_{\mathrm{e}} = 15/2$) D2 transition of the Fr atom, with a wavelength of 718.216 nm. The output power of the Ti:S laser was 3.5 W for a pump laser light output power of 18 W at 532 nm. After passing through a Faraday isolator (FI), part of the laser beam was reflected by a beam splitter (BS) and input into a wavelength meter (WM). The transmitted laser beam was divided into two beams for magneto-optical trapping of the Fr atoms and frequency offset locking by a polarizing beam splitter (PBS). The latter was phase-modulated using a polarization-maintaining fiber (PMF)-coupled EOM (Photline Technologies, NIR-MPX800-LN-05), which was driven by an RF synthesizer with a frequency of 4.6 GHz to generate sidebands in an optical spectrum. The EOM consists of a single-pass waveguide and possesses low driving voltage and large modulation bandwidth \cite{Peng}. The sideband frequencies $\nu_{\pm n}$ generated by the EOM were labeled $\nu_{\mathrm{t}} \pm n \nu_{\mathrm{rf}}$, where $\nu_{\mathrm{t}}$ and $\nu_{\mathrm{rf}}$ are the trapping frequency and RF from the synthesizer, respectively. Further, $\pm n$ indicates the order of the positive or negative sideband ($n = 0, 1, 2, \cdots$, $\nu_0$ corresponds to the trapping frequency), and $(+)n$ indicates a higher frequency than the trapping frequency. The RF power input to the EOM was amplified by an RF amplifier (RFHIC Corporation, RUM43020-10) up to approximately 37.1 dBm, so as to yield the $\pm$10th-order frequency sideband components. The power of incident light into the EOM was 4.5 mW, and the power of transmitted light was typically 0.78 mW (17\% transmission efficiency). Note that the low transmission efficiency is due to the fact that the 718-nm wavelength is outside the specified range of the product \cite{Photline}.

\begin{figure}
\centerline{\includegraphics[width=0.9\columnwidth]{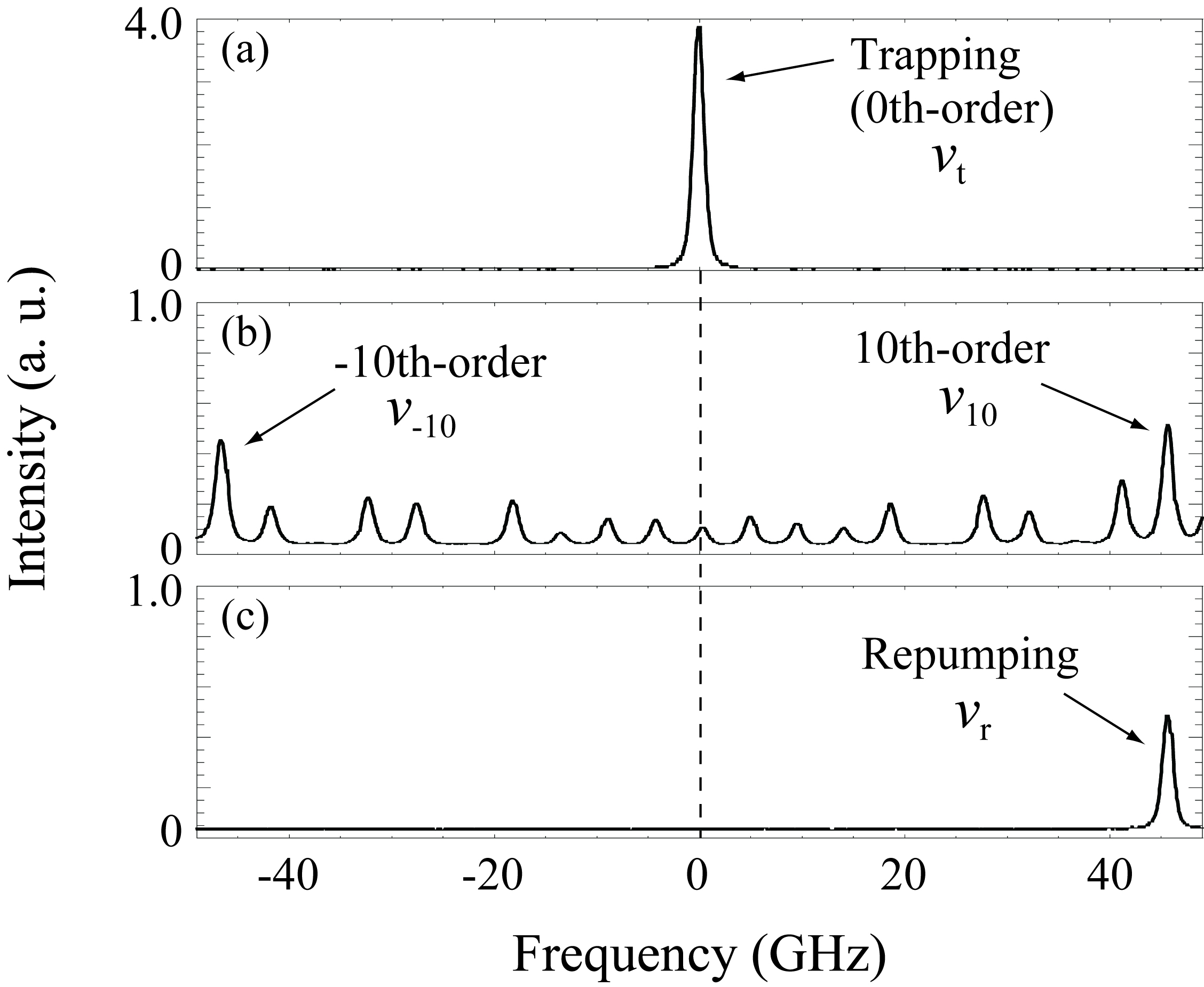}}
\caption{Optical spectra measured by PD1 for (a) the trapping light without EOM modulation, (b) the trapping light modulated by 4.6-GHz RF with 37.1-dBm power, and (c) the repumping light.}
\label{10thsideband}
\end{figure}

Further, we employed a Littrow-type external-cavity laser diode (ECLD, TOPTICA Photonics) as the repumping light source. The wavelength of the light was tuned to the $7S_{1/2}$ ($F_{\mathrm{g}} = 11/2$) $-$ $7P_{3/2}$ ($F_{\mathrm{e}} = 13/2$) transition at 718.137 nm and its output power was 36 mW. The laser beam from the ECLD was split by a PBS after it passed through a FI and a BS, and it was input into a PMF using a fiber coupler (FC).

\begin{figure}
\centerline{\includegraphics[width=0.9\columnwidth]{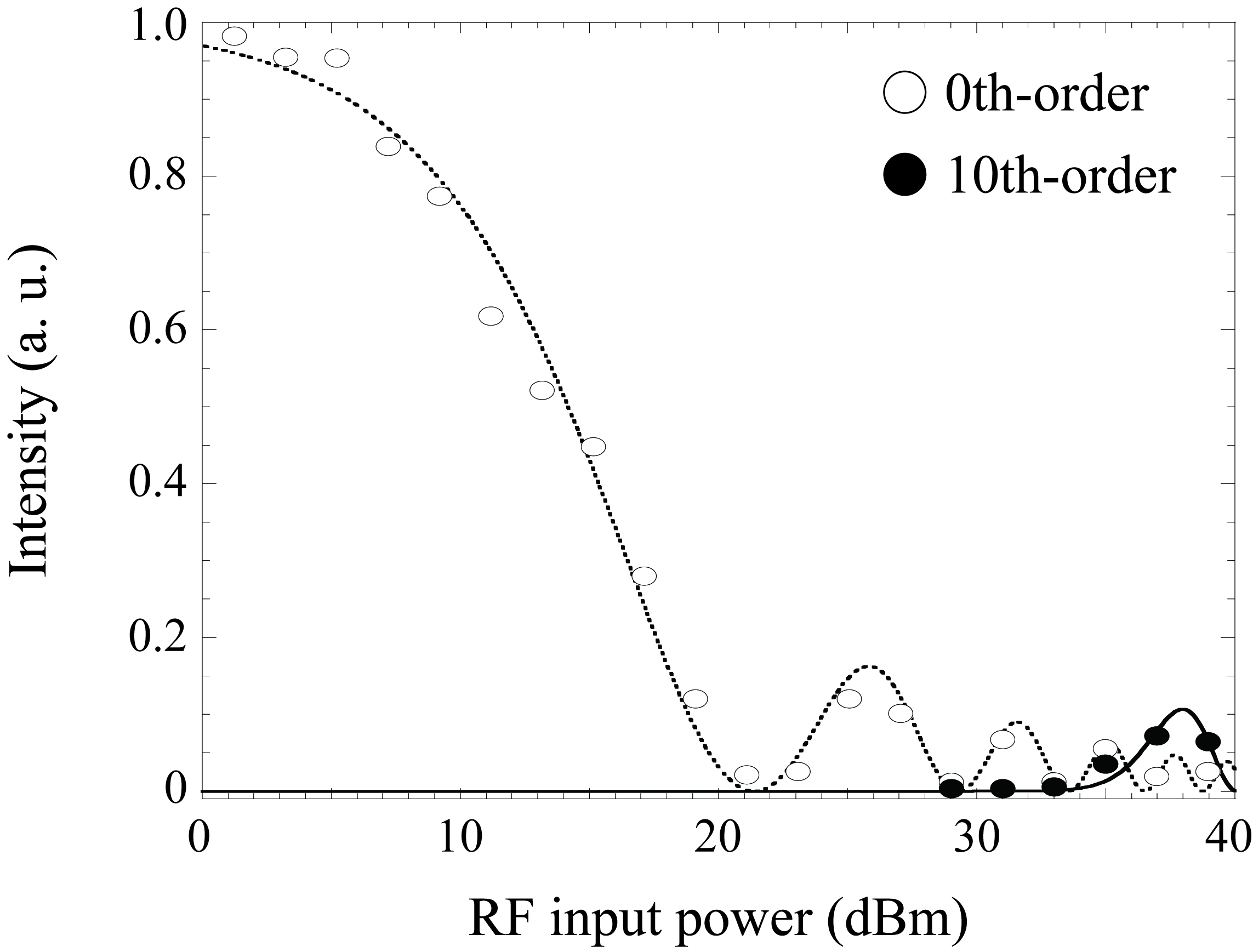}}
\caption{Output intensities of 0th- and 10th-order components as functions of RF input power. The open (filled) circle indicates experimental data for 0th (10th) order. The dashed (solid) line is the 0th (10th)-order Bessel function fit to the data. Both intensities are normalized as the intensity of 0th-order without RF input power is one.}
\label{Amplitude}
\end{figure}

The trapping and repumping lights that is outgoing from the PMFs on output of EOM were divided into two beams and overlapped by a 50:50 non-polarizing beam splitter (NPBS). They were detected by photodetector 1 (PD1) after passing through a cavity "FPC1", so as to monitor their spectra, and by photodetector 2 (PD2, Thorlabs FPD310-FV), so that the beat signal could be observed. The PD2 is useful for detection of laser power modulated with high frequency ranging from 1 MHz to 1.5 GHz. The signal to noise ratio of the beat signal was about 50 dB, and the obtained 3dB linewidth of the beat signal between the carrier and ECLD (the 10th sideband and ECLD) was about 1.0 MHz (1.2 MHz) observed by using a spectrum analyzer with the resolution bandwidth (RBW) of 300 kHz and the sweep time of 10 ms. FPC1 consisted of two mirrors (Layertec, \#110371) with 99\% reflectivity and 12.7-mm diameter. As the distance between the two mirrors was 1.5 mm, the free spectral range of the cavity was estimated to be $\sim$ 100 GHz. The cavity length could be varied using a hollow-piezoelectric actuator. Further, the electric signal output from PD2 was amplified by 20 dB and subsequently mixed with the signal output from a function generator (FG) using a double balanced mixer 1 (DBM1, shown in Fig.2). The polarization of the repumping light was adjusted by a half-wave plate ($\lambda$/2) in order to maximize the beat signal between the two laser lights. As we confirmed the 10th-order sideband only, we consider the beat frequency $\nu_{\mathrm{r}} - \nu_{10}$ in the following. The output frequencies from the DBM1 (shown in Fig.2) were represented by $\Delta_{\pm} = | \left( \nu_{\mathrm{r}} - \nu_{10} \right) \pm \nu_{\mathrm{fg}} |$, where $\nu_{\mathrm{r}}$ and $\nu_{\mathrm{fg}}$ are the repumping and output frequencies from the FG, respectively. The signal was split into two by a power splitter (PS) and one part was delayed by a 1-m-long coax cable (indicated as "Delay cable"), before both signals were recombined by a second DBM (DBM2) again. (Note that the length of shorter path is almost negligible because the PS and DBM2 is directly connected, concerning the port of shorter pass.) The signal mixed by the DBM2 was expressed as \cite{Schunemann}

\begin{align}
\cos &\left( 2 \pi \Delta_{\pm} t \right) \times \cos \{ 2 \pi \Delta_{\pm} \left( t + \tau \right) \} \notag \\
&= \dfrac{1}{2} \left[ \cos \{ 2 \pi \Delta_{\pm} \left( 2t + \tau \right) \} + \cos \left( 2 \pi \Delta_{\pm} \tau \right) \right],
\end{align}
where $\tau$ represents the delay time, which is dependent on the difference in length between the two cable. The obtained output voltage varied as a function of $\cos \left( 2 \pi \Delta_{-} \tau \right)$ \cite{Schunemann} because the higher harmonic components in the equation were suppressed considerably by passing through a 1-MHz low-pass filter (LPF). The error signal was input into a proportional-integral circuit (PIC) and then fed back to the piezoelectric driver for the ECLD, in order to fix the frequency difference between the two light sources.

\section{Results}

\begin{figure}
\centerline{\includegraphics[width=0.9\columnwidth]{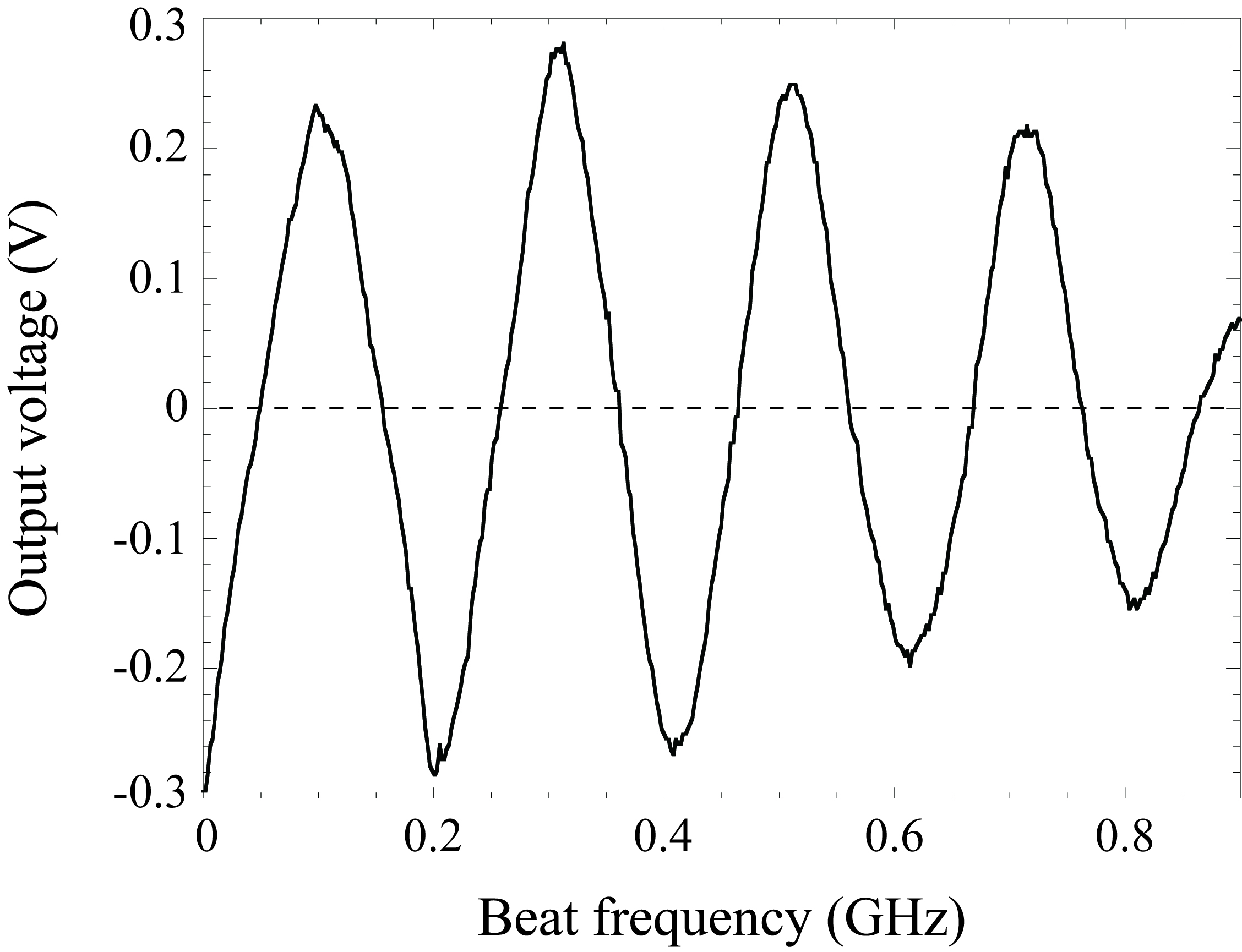}}
\caption{Error signals as functions of beat frequency between 10th-order and repumping components for a 1-m cable length. The attenuated amplitude of the cosine curve was caused by the limited bandwidth of PD2.}
\label{Error}
\end{figure}

To confirm sideband generation for the trapping light after the beam had passed the EOM, we first measured the output powers from FPC1. The results are shown in Fig. \ref{10thsideband}. The frequency, which is shown on the horizontal axis, was obtained by adjusting the cavity length. Its width corresponds approximately to the FSR. Figure \ref{10thsideband}(a) shows the trapping frequency without RF input to the EOM. No sideband components are apparent. In contrast, $\pm$10th-order sideband components appeared with an RF signal input of 37.1 dBm to the EOM, as shown in Fig. \ref{10thsideband}(b). The frequency difference between the neighboring peaks was 4.6 GHz ($\pm$5th and $\pm$8th-order sideband peaks do not appear at this RF power), and the intensities of the sidebands of each order agreed with the corresponding Bessel function. The power ratio of the 10th- to the 0th-order components without modulation was approximately 10\%. The repumping frequency in Fig. \ref{10thsideband}(c) was 46 GHz from the trapping frequency and almost corresponded to the 10th-order sideband component. A beat signal of a few hundred MHz, which was produced by the interference between the two beams, was observed at PD2. The input powers of the trapping and repumping beams to the cavity were 720 and 60 $\mu$W, respectively.

\begin{figure}
\centerline{\includegraphics[width=1\columnwidth]{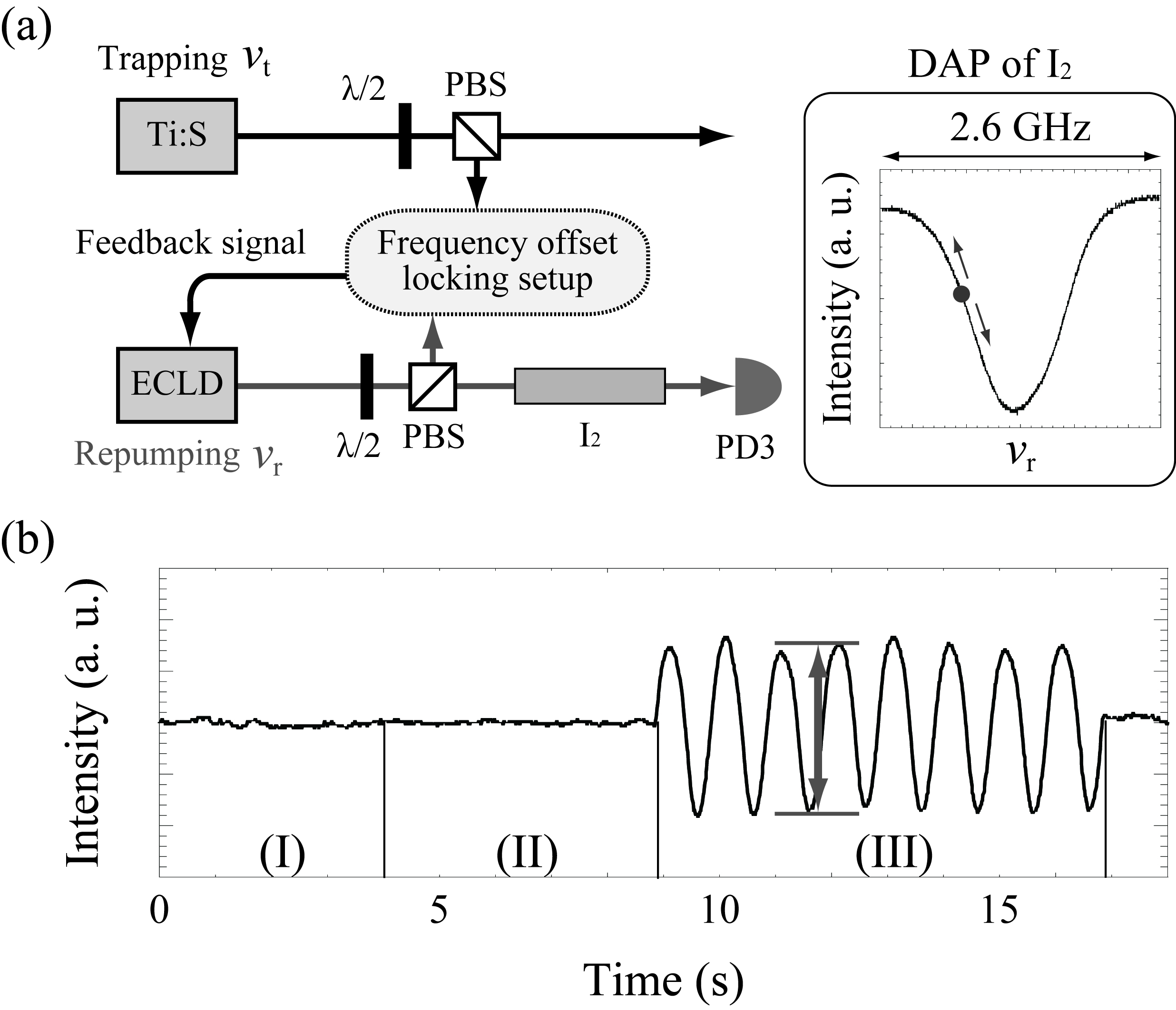}}
\caption{(a) Frequency measurement of repumping light by monitoring the DAP of I$_2$ during the offset locking to the trapping light frequency. The inset shows the DAP obtained by scanning the repumping frequency. The dot in the inset represents the initial position of the repumping frequency. (b) Repumping transmission intensity obtained from DAP slope. (I) Free running. (II) Repumping light frequency locked to trapping light. (III) Repumping frequency is swept by sweeping the frequency of trapping light. An arrow indicates the 1-GHz range.}
\label{Absorption}
\end{figure}

We then observed the peak intensities of the 0th- and 10th-order sideband components of the trapping light as functions of the input RF power, as measured by PD1. The result is shown in Fig. \ref{Amplitude}. When the RF power input to the EOM was in the range from 34 to 40 dBm, the 10th-order components appeared. The signal intensities were in good agreement with the theoretical curves given by 0th- and 10th-order Bessel functions. The intensity ratio of a single 10th-order component to the 0th-order without RF power was estimated to be approximately 10\% at 38 dBm.

The output voltage after passing through the LPF after DBM2 is shown in Fig. \ref{Error}, as a function of beat frequency of $\nu_{\mathrm{r}} - \nu_{10}$. The incident powers of the trapping and repumping lights on PD2 were 0.4 and 1.5 mW, respectively, and the peak spectral power at the 200-MHz beat signal detected by the RF spectrum analyzer was approximately -12.5 dBm. The input signal from the FG was fixed at 100 MHz for the measurement. As the $\tau$ caused by a cable length of 1 m is 5 ns, the frequency of $\cos \left( 2 \pi \Delta_{\pm} \tau \right)$ is 200 MHz, which almost corresponds to 202.7 MHz, which is the observed frequency of periodic signal in Fig. \ref{Error}. This slight difference means that the time difference is shorter by 0.07 ns than 5 ns expected for 1m-cable, which caused from the SMA plug-to-plug connector with length of 1.5 cm on shorter pass between the PS and DBM2. The zero-crossing voltage provides the error signal, which allows for frequency locking using the PIC. In Fig. 5, there are 9 crossing points, which are useful for frequency locking. The zero-crossing positions can be tuned precisely by adjusting the frequency produced by the FG.

To confirm that the frequency difference between the trapping and repumping lights was locked during the frequency of trapping light is sweeping, we then obtained the absorption spectrum of I$_2$. The simple setup used in this experiment is schematically illustrated in Fig. \ref{Absorption}(a). The repumping wavelength was tuned to 718.152 nm, which matches the slope of the Doppler absorption profile (DAP) of the vibrational-rotational transition of I$_2$ \cite{Aime} (see Fig. \ref{Absorption}(a), inset). This absorption line of I$_2$ is close to the Fr repumping transition. As the frequency difference between the 10th-order component of the trapping and repumping lights was fixed by the feedback signal to the ECLD from the PIC, the repumping beam frequency followed the frequency sweep of the trapping light to keep the beat frequency constant. The light from the ECLD passed through an I$_2$ cell and was subsequently transferred to a third photodetector (PD3). The input power was 60 $\mu$W, and 20-cm-long cell was heated to 573 K. When the repumping frequency was positioned at the center of the slope, the frequency change was converted into the amplitude based on the values detected using PD3. Figure \ref{Absorption}(b) shows the transmitted intensity of repumping light, which relates to the repumping frequency via the slope of DAP. After the repumping frequency was offset-locked, the trapping frequency was swept by a triangular waveform voltage with a repetition frequency of 1 Hz, as shown in the range (III), which is from 9 to 17 s on the horizontal axis, in Fig. 6 (b). It was confirmed that the frequency of the repumping light, as monitored by PD3, was also swept by the triangular waveform with 1-Hz repetition frequency. The scanned frequency width was estimated to be approximately 1 GHz based on the DAP voltage, which corresponds to the sweeping width of the trapping frequency. The maximum capture range was 1.4 GHz in the case of the 1 m cable length. This capture range is limited by mode hop of the repumping light frequency because this ECLD is in a Littrow configuration. If one uses a Littman configuration, the capture range will be extended to be more than 30 GHz, which is limited by Ti:Sapphire laser (trapping light source).

\section{Conclusion}

We performed the laser frequency locking with 46 GHz offset using an EOM with 10th-order sidebands for magneto-optical trapping of Fr atoms. The error signal related to the beat frequency between the 10th-order component and the repumping was obtained using the delayed self-homodyne detection technique. Sweeping the trapping-light and repumping-light frequency with keeping its frequency difference of 46 GHz was confirmed over 1.4 GHz by monitoring the Doppler absorption profile of I$_2$. This technique enables us to search for a resonance frequency of magneto-optical trapping of Fr.

\section*{Funding Information}

Grant-in-Aid for Scientific Research (21104005, 26220705);
Murata Science Foundation;
One of the authors (Y. Sakemi) is supported by JSPS and INSA under the Japan-India Research Cooperative Program.


\begin{thebibliography}{99}
\bibitem{Photline}
The NIR-MPX800-LN-05 is especially designed to operate in the 780-890-nm wavelength.
\end{thebibliography}
\end{document}